\documentstyle[prl,twocolumn,aps]{revtex}
\newcommand{\lapx}{\mbox{\raisebox{-4pt}{$\,\buildrel<\over\sim\,$}}}

\begin{document}
\draft
\title{Bulk and boundary zero-bias anomaly in multi-wall carbon nanotubes}
\author{R. Egger$^{1,2}$ and A.O. Gogolin$^1$}
\address{${}^{1}$ Department of Mathematics, Imperial College, 180 Queen's Gate,
London SW7 2BZ, United Kingdom\\
${}^{2}$ Fakult\"at f\"ur Physik, Albert-Ludwigs-Universit\"at 
Freiburg, D-79104 Freiburg, Germany
}
\date{Date: \today}
\maketitle
\begin{abstract}
We compute the tunneling density of states of doped multi-wall 
nanotubes including disorder and electron-electron interactions.
A non-conventional Coulomb blockade
reflecting nonperturbative Altshuler-Aronov-Lee power-law zero-bias
anomalies is found, in accordance with recent experimental results.
The presence of a boundary implies 
a universal doubling of the boundary exponent in the diffusive limit. 
\end{abstract}

\narrowtext

Carbon nanotubes provide a remarkable and exciting arena for
mesoscopic transport phenomena involving strong 
electron correlations \cite{dekker99}.  
Recent experiments on single-wall nanotubes (SWNTs) have 
established the ballistic nature of SWNT transport and revealed the
Luttinger liquid (LL) behavior 
of 1D interacting fermions \cite{ll1}. 
However, the situation is less clear for
multi-wall nanotubes (MWNTs), which are composed
of several (about ten) concentrically arranged
graphite shells, where experiments 
seem more consistent with diffusive transport \cite{sc1},
e.g.~showing typical weak localization features in
the magnetoconductance.  

Quite remarkably, several experimental observations for MWNTs 
do not seem to fit into the framework of existing theories.  
The most prominent example concerns the pronounced power-law
zero bias anomaly of the tunneling density of
states (TDOS) found at low energy scales, $E\lapx 0.1$~eV 
\cite{sc2,sc3},
\begin{equation} \label{obs}
\nu(E)\sim E^\alpha \;, \quad 
\alpha\approx 0.3 \;.
\end{equation}
Furthermore, the TDOS at the end of the MWNT, while still
of power-law form, is characterized by the
boundary exponent $\alpha_{\rm end}\approx 0.6$. 
In this Letter, we show that these results can be understood
in terms of a particularly effective
and non-conventional {\sl Coulomb blockade} (CB) 
for tunneling into a strongly interacting disordered metal.
Charge propagation on the MWNT is then effectively two-dimensional (2D), 
but for very low energy scales not probed in Ref.\cite{sc3},
$E<E^*$ with $E^*$ in Eq.~(\ref{estar}) below,
a crossover to 1D pseudo-gap behavior \cite{gl1} is expected. 

The key ingredient in CB theory is the probability $P(E)$ that 
a tunneling electron excites electromagnetic modes
 with energy $E$ in the system \cite{sct}.  The theory
is meaningful if these modes are harmonic, and then $P(E)$
directly determines the TDOS,
\begin{equation} \label{tdossct}
\frac{\nu(E)}{\nu_0} = \int_{-\infty}^\infty dE' \frac{1+\exp(-E/k_B T)}{1+
\exp(-E'/k_B T)} P(E-E') \;,
\end{equation} 
where $\nu_0$ is the non-interacting DOS.
The probability $P(E)$ is the Fourier transform of 
$\exp [J(t)]$,  with the phase correlation function
($\hbar=1$)
\begin{eqnarray} \nonumber 
J(t) & = & \int_0^\infty d\omega \frac{I(\omega)}{\omega}
\Bigl \{\coth(\omega/2 k_B T) [\cos(\omega t)-1] \\
&& \quad \quad - i \sin(\omega t) \Bigr \} 
\label{jf}
\end{eqnarray}
for a spectral density $I(\omega)$ of electromagnetic modes.
For simplicity, we now focus on the zero-temperature case.
Provided $I(\omega)$ remains finite for low frequencies,
Eqs.~(\ref{tdossct}) and (\ref{jf}) then straightforwardly lead to
the power law (\ref{obs}) with exponent $\alpha = I(\omega\to 0)$.

In a conventional CB system, $I(\omega)$ is phenomenologically 
parameterized in terms of the total impedance $Z(\omega)$ \cite{sct},
and one obtains $\alpha= Z(0)/(h/2e^2)$.
Such a transmission line model directly explains
the doubling of the end exponent, as in the bulk
case one has effectively two resistances in parallel as
compared to the end case.  Applying this model to MWNTs, however, 
one finds rather small values for $\alpha$, and, in addition,
the observed voltage dependence cannot be reproduced \cite{sc3}.
On the other hand, for a LL, the elementary excitations are
harmonic, $I(\omega)$ is known, and hence $P(E)$ can also be
computed exactly.  The resulting exponents can be written in terms of  
the standard interaction parameter $g$,  and the ratio
$\alpha^{\rm LL}_{\rm end}/\alpha^{\rm LL}$ is
$g$-dependent and always larger than two \cite{kf,foot}.
In MWNTs $g$ is modified by inner-shell screening, and 
the relation between $\alpha^{\rm LL}$ and $g$ is 
affected by the available subbands \cite{egger99}.
Estimating $\alpha^{\rm LL}$ and $\alpha_{\rm end}^{\rm LL}$ 
for the situation of Ref.\cite{sc3},
they are at least one order of magnitude smaller than observed.
Hence neither conventional CB nor LL theory can explain these data. 

For a ballistic system, the full crossover in the CB between a 
(single-channel) LL and a (many-channel) classical resistor has been
worked out in Ref.\cite{matveev}.  The results of this
study imply that it is crucial to take disorder into account here.
We thus have to (i) establish the harmonic nature
of the electromagnetic modes, (ii) compute the low-frequency
spectrum $I(\omega)$ and hence the exponent $\alpha$,
and (iii) compute the end exponent.  
Note that a perturbative treatment of the
interactions is not sufficient, as
the power law (\ref{obs}) is inconsistent
with conventional (1D or 2D) Altshuler-Aronov-Lee (AAL)
predictions for the TDOS \cite{altshuler}. However, 2D AAL logarithmic
dependencies have been observed for MWNT bundles,
which presumably are characterized by weaker interactions \cite{Lu}.  

The main difference between MWNTs and SWNTs, apart from
the larger radii of MWNTs, $R\approx$ 4 to 10 nm,
stems from the presence of inner shells.
Although inter-shell tunneling is largely suppressed for 
a number of reasons \cite{kane},
and hence transport proceeds only through the outermost contacted shell
\cite{sc1}, inner shells cause a screening of the electron-electron
interaction.  For a computation of the TDOS, the latter effect, as well as 
spin and Fermi (K) point degeneracy \cite{dekker99}, 
can be included by a renormalization of the 
electron-electron interaction potential $U(\vec{q})$ \cite{egger99},
and hence we consider spinless electrons with 
only one K point in what follows.
Since different shells always have incommensurate lattices,
a {\sl quasiperiodic} ionic potential from inner shells acts on
outermost-shell electrons.  The effect of such a potential
is very similar to a random potential described by a 
mean free path $\ell=v\tau$ \cite{bak},
which can be estimated from a disordered
tight-binding approach using random on-site energies
with variance $\sigma_E^2$. For nearest-neighbor
hopping $t_0$ and $2N+1$ subbands (see below), one finds \cite{egger99}
\begin{equation} \label{mfp}
\ell = (t_0/\sigma_E)^2 R/(2N+1) \approx 10 R\;.
\end{equation}
Here $\sigma_E\approx t_0/10$ corresponds to the hopping strength between
adjacent shells, and $N\approx 5$.
Therefore disorder is present even in impurity-free MWNTs. 
In addition, ``true'' disorder imposed by imperfections, substrate
inhomogeneities, or defects can be important.
Using Matthiessen's rule,
$\ell^{-1}\to \ell^{-1} + \ell_0^{-1}$,
where $\ell_0$ is the mean free path due to ``true'' disorder.
Typical values of $\ell \approx 5$ to 100~nm
were reported in Refs.~\cite{sc1,sc2,sc3}.
Hence disorder is still ``weak'', $k_F \ell \gg 1$,
and the possibility of a diffusive phase of interacting electrons
must be discussed.  Doped MWNTs are in between the 1D and  
2D limits and may be expected to show diffusive behavior 
over a wide energy range, 
where the Anderson localization expected for truly 1D disordered systems 
has not yet set in.  We mention in passing that the phase
relaxation length is $\ell_\phi\approx 500$~nm at 2~K, with
the usual $T$ dependence due to electron-electron
interactions \cite{sc2}.

The bandstructure of a clean nanotube is described by a
Dirac ``light cone'', $E(\vec{k})=v|\vec{k}|$,
with $\vec{k}=(k,k_\perp)$ and
quantized transverse momentum, $k_\perp=n/R$,
where $n=-N,\ldots,N$ and $N=[k_F R]$.  The Fermi
velocity is $v=8\times 10^5$ m/sec, and
the number $2N+1$ of 1D subbands arising from periodic
boundary conditions around the circumference is 
determined by the doping level $\mu$ via $k_F=\mu/v$. 
Since MWNTs studied experimentally so far
are characterized by rather large doping, $\mu\approx 0.5$~eV 
\cite{sc4}, a typical value is $N\approx 5$.
The $n$th subband is characterized by Fermi velocity and momentum,
\begin{equation}
v_n=v \sqrt{1-(n/k_F R)^2} 
\;, \quad k_n = k_F v_n/v \;.
\end{equation}
The theoretical description for small $\mu$ is
quite intricate and will be given elsewhere, 
as the disorder causes logarithmic divergences 
of the self energy due to the vanishing DOS of 
the Dirac cone \cite{ntw}.  
Fortunately, these complications are absent for the case
studied below, $k_F R \gg 1$.  In the following,
we focus on energy scales $E < v/R$, where
it is sufficient to take a fixed $N$ and thereby ignore
van Hove singularities associated with the opening 
of new 1D bands as the energy varies.
 
Let us start with the  bulk TDOS, where  
standard diagrammatic perturbation theory \cite{altshuler}
adopted to the MWNT geometry is quite illuminating. 
 The disorder-averaged Greens function is 
\begin{equation}\label{g0}
G_{R/A}(E,\vec{k}) = [E+\mu-E(\vec{k}) + i
\, {\rm sgn} (E)/2\tau ]^{-1} \;,
\end{equation}
and hence the non-interacting DOS is 
\begin{equation}\label{dos}
\nu_0 = - \frac{{\rm Im}}{\pi} \int \frac{dk}{2\pi} \sum_{k_\perp} 
G_R(E,\vec{k}) = \sum_{n=-N}^N \frac{1}{\pi v_n} \;.
\end{equation}
Prefactors in expressions like Eq.~(\ref{dos}) are 
always chosen to ensure the correct 1D limit
for $N=0$. In the 2D limit, $N\gg 1$, the above
DOS is related to the conventional 2D DOS, $\nu_{\rm 2D}=k_F/2\pi v$,
via $\nu_0=2\pi R \nu_{\rm 2D}$.  
The vertex renormalization due to one impurity line 
is described by
\begin{eqnarray} \nonumber
\xi(\omega, \vec{q}) &=& \frac{1}{\pi \nu_0 \tau} \int\frac{dk}{2\pi}
\sum_{k_\perp} G_R(E+\omega,\vec{k}+\vec{q}) \, G_A(E,\vec{k})
 \\  \label{xii}
&=& \sum_{n} \sum_{\sigma=\pm}
 \frac{1}{ 2\pi \nu_0 v_n} \Bigl[ 1- i\omega \tau \\
\nonumber && +i\sigma v_n\tau \left(q+
 n q_\perp/\sqrt{ (k_F R)^2 - n^2}\right) \Bigr]^{-1} \;,
\end{eqnarray}
where $q_\perp=m/R$ is also quantized.
In the low-energy long-wavelength limit, $\omega\tau,
|\vec{q}|\ell \ll 1$,  Eq.~(\ref{xii}) implies
\begin{equation} \label{xi2}
\xi(\omega, \vec{q}) = 1+i\omega\tau - 
D q^2 \tau - D_\perp q_\perp^2 \tau \;,
\end{equation}
with diffusion constants $D = (\tau/\pi\nu_0) \sum_n v_n$ and
\[
D_\perp = (\tau/\pi\nu_0) (v/k_F R)^2 \sum_n n^2 / v_n \leq D \;.
\]
One checks easily that $D_\perp \to D$ for large $k_F R$, 
with $D\to v^2 \tau/2$.  For $N=0$, however, $D_\perp=0$ and
$D=v^2\tau$, and therefore both 1D and 2D limits are correctly
reproduced.  Summing the usual ladder series then directly
implies a diffusion pole, 
$[1-\xi(\omega,\vec{q})]^{-1}$. 

To ensure the validity of the ladder approximation,
however, the smallness of ``crossed'' diagrams first
needs to be checked.  Since here one is actually in between the 1D
and 2D limits, the usual reasoning leading to a suppression factor 
$1/k_F \ell$ for crossed diagrams does not apply.
Evaluation of the simplest diagram of the crossed type 
and comparing it to the corresponding ladder diagram gives
a suppression factor
$f = (1/2) \sum_n 1/(\pi \nu_0 v_n)^2$.
Since $f=1/2$ for $N=0$, it is clearly not justified to neglect crossed
diagrams for small doping level $\mu$ where one approaches the 1D limit.
The resulting breakdown of the diffusive picture in this limit 
is of course expected from 1D localization theory.  
For $k_F R\gg 1$, however, we obtain $f \simeq\ln(k_F R)/[2\pi^2 k_F R]
\ll 1$, and are thus entitled to neglect crossed diagrams. 
In that case, $D_\perp\simeq D=v^2 \tau/2$.  

Here we consider an effectively short-ranged 
1D interaction, $U(\vec{q})\approx U_0$,
since the $1/r$ tail of the Coulomb interaction potential
is often externally screened, e.g.~by close-by metallic gates.
Even when working on an insulating substrate, a situation characterized by
a long-ranged interaction, the finite length 
of the MWNT will lead to a cutoff for the logarithmic
singularity.  Note that $U_0$ represents a 1D Fourier component 
and hence is appropriate for screening lengths $\lambda_s$
large compared to $R$.   
In the 2D limit, $R\gg \lambda_s$,
this interaction constant is related 
to the respective 2D parameter via $U_0= U_{\rm 2D}/R$.
For the case of interest here, $R\ll \lambda_s$, and the order-of-magnitude
estimate $U_0/2\pi v \approx 1$
obtains \cite{egger99}, implying a dielectric constant
$\epsilon\approx 10$.  Under an extended Hubbard model description,
this amounts to $U/t\approx$ 1 to 5 in standard terminology.  
The first-order interaction
correction to the TDOS can then be obtained 
following well-known arguments \cite{altshuler,glazman}.
Since the Hartree correction is subleading, we focus on the 
exchange correction, 
\begin{equation} \label{tdos1}
\frac{\delta \nu(E)}{\nu_0} = -\frac{\tau}{\pi} {\rm Re}
\int \frac{dq}{2\pi} \sum_{q_\perp} U(\vec{q}) 
\frac{\xi(E,\vec{q})}{1-\xi(E,\vec{q})} \;.
\end{equation}
Together with Eq.~(\ref{xii}), this result
describes the complete crossover from the ballistic LL 
to the diffusive limit.
For $\tau\to \infty$, no energy dependence is found, in
agreement with $\alpha^{\rm LL}=0$ to this order in $U_0$.  
In the diffusive limit, however, Eq.~(\ref{tdos1}) can be simplified to 
\begin{equation}\label{xxx}
\frac{\delta \nu(E)}{\nu_0} = - \frac{U_0}{2\pi \sqrt{D}} {\rm Re}
\sum_{n=-N}^N \left(  -i E + D n^2/R^2  \right)^{-1/2}\;.
\end{equation}
The $n=0$ contribution yields the $E^{-1/2}$ 
1D AAL correction, while the $n\neq 0$ contributions
give a 2D logarithmic correction \cite{altshuler,glazman},
\begin{equation} \label{aal1}
\delta \nu(E)/\nu_0= \alpha \ln(E/E_0) \;,
\end{equation}
with $E_0 \approx\mu$ and $\alpha  = U_0 R /(2\pi D)$. 
In fact, Eq.~(\ref{aal1}) also holds in the quasi-ballistic
regime, $E\tau\gg 1$ \cite{glazman}.
Comparing the 1D and 2D contributions in Eq.~(\ref{xxx}), 
we see that only for sufficiently low energy scales, 
$E < E^\star$, the 1D AAL law, $\delta\nu\sim E^{-1/2}$, takes over, where
we estimate
\begin{equation} \label{estar}
E^\star \approx \frac{v \ell}{16 R^2 
\ln^2( k_F R ) } \lapx 1 \; {\rm meV} 
\end{equation}
for the crossover scale.  Equation (\ref{estar}) holds for
weak-to-intermediate interactions, but may change for
very strong interactions.  For $E > E^\star$, Eq.~(\ref{aal1}) then
gives the leading contribution, and the TDOS 
should be essentially equivalent to the one of a 2D 
disordered interacting metal.  
 
Since in MWNTs, interactions are of intermediate strength, 
the lowest-order result (\ref{aal1}) is not sufficient.
Higher-order terms in $U_0$ were recently discussed
by Kamenev and Andreev \cite{ka}. In the framework of the dynamical Keldysh
formalism for a 2D system with long-ranged Coulomb interactions,
a derivation of the TDOS valid for arbitrary interaction
strength has been presented.  Within the saddle-point 
approximation for the emerging nonlinear $\sigma$  model, 
the electromagnetic modes are indeed Gaussian, 
with spectral density  \cite{ka}
\begin{equation} \label{iw} 
I(\omega) = \frac{\omega}{\pi} {\rm Im}
\sum_{\vec{q}} \frac{1}{(Dq^2 - i\omega)^2} \left( U_0^{-1}+
\frac{\nu_0 Dq^2}{Dq^2-i\omega}\right)^{-1} \;.
\end{equation}
Adopting this result to MWNTs in the energy regime $E > E^*$,
the bulk exponent $\alpha=I(\omega\to 0)$  follows as
\begin{equation}
\alpha = \frac{U_0 R}{2\pi D} \, \frac{\ln(1+\nu_0 U_0)}{\nu_0 U_0} \;.
\end{equation}
For weak interactions, the second factor is close to unity, and
thus $\alpha$ is just the prefactor in Eq.~(\ref{aal1}).
In effect, the AAL logarithmic correction therefore {\sl exponentiates}
into a power law.  Using $U_0/2\pi v\approx 1$ and $N\approx 5$, 
we then estimate $\alpha\approx R/\ell$. To compare with experiment,
note that $\ell\approx 10 R$ from Eq.~(\ref{mfp}), so long as only
intrinsic disorder via the inner-shell potentials is present.
Then one gets the same exponent for different tubes,  $\alpha\approx 0.1$. 
When ``true'' disorder is included,  $\alpha$
should increase and depend on the particular MWNT.
This appears to be in qualitative agreement with Ref.\cite{sc3},
where values between $\alpha=0.24$ and $0.37$
were observed for 11 different MWNTs.

Next we turn to the {\sl boundary TDOS}\, by considering 
a semi-infinite MWNT, $x\geq 0$. 
In the absence of disorder, the boundary Greens function is ($\delta=0^+$)
\[
G_0^b(E, k_\perp; x,x') = \int_0^\infty \frac{dk}{2\pi}
\frac{4\sin(kx) \sin(kx')}{E+\mu-E(\vec{k}) +i \delta 
\,{\rm sgn}(E)} \;,
\]
which can be written as $G_0^h(x-x')-G_0^h(x+x')$ with the
homogeneous Greens function $G_0^h$.  Therefore the boundary 
simply causes the ``image term'' $-G_0^h(x+x')$. 
To include disorder, it is convenient to solve the 
Dyson equation in a real-space formalism.
In the Born approximation, the self energy is 
$\Sigma(x) = (\pi \nu_0 \tau)^{-1} [ G_0^h(0) - G_0^h(2x) ]$. 
If we retain only the $G_0^h(0)$ term in the self energy,
the disorder-averaged boundary Greens function  is
\begin{equation}\label{g1}
G^b(E, k_\perp; x,x') = \int_0^\infty \frac{dk}{2\pi}\,
4\sin(kx) \sin(kx')  G(E,\vec{k}) \;,
\end{equation}
where we use Eq.~(\ref{g0}).  In the 1D limit, $k_F R\lapx 1$, the
image part $-G_0^h(2x)$ in the self energy
 is crucial and modifies the structure of $G^b$.
For the more interesting limit
$k_F R\gg 1$, it suffices to analyze a semi-infinite plane,
with the lowest-order correction from the image part 
\begin{eqnarray}\nonumber
\delta G^b(\vec{r},X)&=&-\frac{1}{\pi\nu_{\rm 2D}\tau}\int d\vec{x}''
G^h_0\left(\frac{\vec{r}}{2}-\vec{x}''\right)\\ &\times& 
G^h_0(2x''+2X,0)G^h_0\left(\frac{\vec{r}}{2}+\vec{x}''\right)\;,
\label{impart}
\end{eqnarray}
where the integration extends over the half plane,
$X=(x+x')/2$ is the distance to the boundary, and 
$\vec{r}=\vec{x}-\vec{x}'$ is the relative coordinate.
For $\vec{r}=0$, this can be seen to vanish identically.
Under the conditions that $k_FX\gg 1$ and $k_Fr\gg 1$, Eq.~(\ref{impart})
 simplifies to
\[
\delta G^b(\vec{r},X)\simeq -\frac{1}{k_F \ell \nu_{\rm 2D}}
G^h_0(2X)G^h_0(r)\;.
\]
This gives subleading corrections to Eq.~(\ref{g1})
that are suppressed by a factor $1/k_FX$ or $1/k_Fr$, whichever
is smaller.  As these corrections die out on the 
microscopic lengthscale $1/k_F$,
we take Eq.~(\ref{g1}) in what follows. 

The real-space diagrammatic perturbation theory can 
then be carried out explicitly.  
Dressing the interaction vertices with impurity lines
within the ladder approximation,
the first-order correction to the TDOS at position $x$ in the
diffusive limit is 
\[
\frac{\delta \nu(E,x)}{\nu_0} = -\frac{\tau}{\pi} {\rm Re}
\int \frac{dq}{2\pi} 2 \cos^2(qx)  \sum_{q_\perp} U(\vec{q})  
\frac{\xi(E,\vec{q})}{1-\xi(E,\vec{q})} \;,
\]
with $\xi(E,\vec{q})$ given by Eq.~(\ref{xi2}).
For $x\gg (D/E)^{1/2}$, we recover the bulk result
(\ref{tdos1}). However, when tunneling into the
end of the MWNT, $x\ll (D/E)^{1/2}$, the prefactor 
in Eq.~(\ref{aal1}) gets {\sl doubled}.
This doubling can be rationalized
using the quasi-classical picture of Ref.~\cite{larkin},
where the TDOS at position $x$ is related to the
Fourier transform of the return probability $P(x,t)$.
The latter satisfies the diffusion equation with the boundary 
condition that no current flows across the boundary, 
 leading to $P(x,t)/P(\infty,t)=1+\exp(-x^2/Dt)$.
Close to the boundary, the return probability is then doubled.   

Repeating the above arguments leading to the power law
(\ref{obs}) for the bulk TDOS then implies again a power law,
now characterized by the boundary exponent
$\alpha_{\rm end} = 2 \alpha$.
The doubling of the exponent as one goes from the bulk to
the boundary is in agreement with the experimental results
of Ref.~\cite{sc3}.  It should be noted that our
derivation holds only in the diffusive limit.
Remarkably, this doubling -- which is based on the properties of
the boundary diffuson close to the edge --
coincides with the prediction of the classical resistor model.

Let us conclude by listing several open questions
raised by our study.  Future work needs to address the 
situation at small doping levels, 
the magnetic field dependence of the TDOS,
and the conductivity.  We hope  that 
future theoretical and experimental work
continues to reveal the subtle and beautiful interplay
of disorder, dimensionality, and interactions presented
by multi-wall nanotubes.

We thank L.I. Glazman for discussions, and A. Bachtold,
Li Lu, and C. Sch{\"o}nenberger for motivating this work
and providing unpublished data.  Support by the 
by the DFG (Gerhard-Hess program) and by the
EPSRC (Grant GR/N19359) is acknowledged.


\begin{references}
\bibitem{foot}
As a consequence, a transmission line
interpretation of the SWNT experiments \cite{ll1}
can be excluded.
\end{references}
\end{document}